\documentclass[12pt]{article}
\usepackage{epsf}
\def\be{\begin{equation}}
\def\ee{\end{equation}}
\def\bea{\begin{eqnarray}}
\def\eea{\end{eqnarray}}

\def\yp{\Upsilon}

\begin{document}
\begin{titlepage}
\begin{center}
{\Large \bf William I. Fine Theoretical Physics Institute \\
University of Minnesota \\}
\end{center}
\vspace{0.2in}
\begin{flushright}
FTPI-MINN-12/25 \\
UMN-TH-3113/12 \\
July 2012 \\
\end{flushright}
\vspace{0.3in}
\begin{center}
{\Large \bf Remarks on decays of $h_b(2P)$
\\}
\vspace{0.2in}
{\bf Xin Li$^a$  and M.B. Voloshin$^{a,b,c}$  \\ }
$^a$School of Physics and Astronomy, University of Minnesota, Minneapolis, MN 55455, USA \\
$^b$William I. Fine Theoretical Physics Institute, University of
Minnesota,\\ Minneapolis, MN 55455, USA \\
$^c$Institute of Theoretical and Experimental Physics, Moscow, 117218, Russia
\\[0.2in]

\end{center}

\vspace{0.2in}

\begin{abstract}
We consider hadronic transitions from the $h_b(2P)$ bottomonium resonance to lower states of bottomonium with emission of either $\omega$ meson, or two pions, or $\eta$ meson. For the former two transitions the branching ratios are related to similar transitions from $\chi_{b1}(2P)$ and the recently measured by Belle fractions of the radiative decays of $h_b(2P)$. We argue that the fraction of the $h_b(2P)$ total decay rate remaining for the annihilation rate is on the verge of contradiction with the `parton' picture of bottomonium annihilation resulting in similarity between the decays of $h_b(1,2P)$ and $\chi_{b1}(1,2P)$. The contradiction gets even stronger, if the transition $h_b(2P) \to \Upsilon(1S) \, \eta$ has branching fraction of a few percent or more. We argue that, although quite uncertain, the latter fraction may indeed be that significant.
\end{abstract}
\end{titlepage}

The spin-singlet $P$ wave $^1P_1$ states of bottomonium $h_b(1P)$ and $h_b(2P)$ provide ample opportunities to study the QCD dynamics of heavy quarkonium. These resonances were observed by Belle~\cite{belleh} in two-pion transitions from $\yp(5S)$ and also an evidence of the transition $\yp(3S) \to h_b(1P) \, \pi^0$ was presented by BaBar~\cite{babar}. The theoretical expectations for the masses and decay properties of these resonances were previously considered in the literature, and the most detailed compilation and discussion can be found in Ref.~\cite{gr}. The theoretical treatment of the $^1P_1$ states is facilitated by their relation, within the nonrelativistic description of bottomonium, to the spin-triplet $\chi_{bJ}(1P)$ and $\chi_{bJ}(2P)$ states. The expected masses of the $h_b$ resonances are determined by the `center of gravity' of the corresponding triplet states (and are in a remarkable agreement with the measured values), while the decay properties of the $h_b(1P)$ and $h_b(2P)$ particles are most naturally related to those of the corresponding $^3P_1$ bottomonium states $\chi_{b1}(1P)$ and $\chi_{b1}(2P)$. Indeed, these latter $J^{PC}=1^{++}$ resonances are the closest in mass to their spin-singlet counterparts, so that the kinematical differences in some decays are minimal, and also they share the property of having relatively small annihilation rates, since in the inclusive `parton' picture of the annihilation both types of states annihilate in the order $\alpha_s^3$: $\chi_{b1} \to q \bar q g$ and $h_b \to 3g$, with $g$ standing for gluon and $q$ for a light quark. The straightforward theoretical picture with a similarity between the decay properties of the $^1P_1$ and $^3P_1$ has been recently put to test by the Belle data~\cite{bellehg} on the dominant radiative transitions from the $h_b$ resonances. The reported branching fraction for such decay of the $h_b(1P)$ resonance, ${\cal B}[h_b(1P) \to \eta_b(1S) \, \gamma] = (49.2 \pm 5.7^{+5.6}_{-3.3})\%$  arguably compares reasonably well (accounting also for a difference in the photon energy) with the known similar fraction for the $\chi_{b1}(1P)$~\cite{pdg}: ${\cal B}[\chi_{b1}(1P) \to \yp(1S) \, \gamma] = (33.9 \pm 2.2)\%$. However, the central values of the data~\cite{bellehg} for the transitions form $h_b(2P)$:
${\cal B}[h_b(2P) \to \eta_b(1S) \, \gamma] = (22.3 \pm 3.8^{+3.1}_{-3.3} )\%$ and  ${\cal B}[h_b(2P) \to \eta_b(2S) \, \gamma] = (47.5 \pm 10.5^{+6.8}_{-7.7} )\%$ are significantly higher than for the spin-triplet `analog' $\chi_{b1}(2P)$~\cite{pdg}: ${\cal B}[\chi_{b1}(2P) \to \yp(1S) \, \gamma] = (9.2 \pm 0.8)\%$ and ${\cal B}[\chi_{b1}(2P) \to \yp(2S) \, \gamma] = (19.9 \pm 1.9)\%$.
Indeed,  the reported central values of the radiative decay rates for $h_b(2P)$ indicate that the annihilation decay rate $\Gamma_{ann}[h_b(2P)]$ of the $h_b(2P)$ may be significantly suppressed in comparison with the rate expected from the similarity relation $\Gamma_{ann}[h_b(2P)] / \Gamma_{ann}[h_b(1P)] = \Gamma_{ann}[\chi_{b1}(2P)]/ \Gamma_{ann}[\chi_{b1}(1P)]$. In this paper we quantify the possible contradiction  with the similarity and argue that it gets even stronger, if the hadronic transitions from $h_b(2P)$ to lower bottomonium states are taken into account. Namely, the $h_b(2P)$ resonance has the decay mode $h_b(2P) \to \eta \, \yp(1S)$, which is  kinematically forbidden for the $h_b(1P)$ and has no heavy-quark-spin analog for the $\chi_{b1}(2P)$ state. We argue that the branching fraction for this decay, although uncertain, can be significant (up to $O(10\%)$), which would further reduce the fractional probability remaining for the annihilation decays of $h_b(2P)$.

The absolute rates of the transitions between the $P$- and $S$-wave states of the spin-singlet bottomonium are related to those for the spin-triplet states by heavy quark spin symmetry within the nonrelativistic description of the $b \bar b$ system. The rates of the radiative electric-dipole  transitions are related as
\be
\Gamma[h_b(kP) \to \eta_b(nS) \, \gamma] = {\omega_{kn1}^3 \over \omega_{kn3}^3} \, \Gamma[\chi_{b1}(kP) \to \yp(nS) \, \gamma]~,
\label{gkn13}
\ee
where $\omega_{kn1}$ ($\omega_{kn3}$) is the photon energy in the transition between the spin-singlet (spin-triplet) states.~\footnote{In a strict sense, the account for the difference in the factor $\omega^3$ is beyond the accuracy of the lowest order in the breaking of the heavy quark spin symmetry. However we follow the tradition of including this factor, since this factor is dictated by the QED gauge invariance, and since the effect of the spin-dependence is somewhat enhanced in this factor. Our conclusions would not change qualitatively, if this factor is omitted.} Using Eq.(\ref{gkn13}) one readily estimates
\bea
&&\Gamma[h_b(1P) \to \eta_b(1S) \, \gamma] \approx 1.5 \, \Gamma [\chi_{b1}(1P) \to \yp(1S) \, \gamma]; \nonumber \\
&&\Gamma[h_b(2P) \to \eta_b(1S) \, \gamma] \approx 1.25 \, \Gamma [\chi_{b1}(2P) \to \yp(1S) \, \gamma]; \nonumber \\
&&\Gamma[h_b(2P) \to \eta_b(2S) \, \gamma] \approx  1.44 \, \Gamma [\chi_{b1}(2P) \to \yp(2S) \, \gamma]~.
\label{gg}
\eea

Unlike the $1P$ states, the heavier $2P$ ones also undergo hadronic transitions to lower levels of bottomonium. Two types  of such transitions, with emission of either $\omega$ resonance or two pions, are common for $\chi_{b1}(2P)$ and $h_b(2P)$ and their absolute rates can be related:
\be
\Gamma[h_b(2P) \to \eta_b(1S) \, \omega]={p_{\omega 1} \over p_{\omega 3}} \, \Gamma[\chi_{b1}(2P) \to \yp(1S) \, \omega] \approx 2.6 \, \Gamma[\chi_{b1}(2P) \to \yp(1S) \, \omega]~,  
\label{go}
\ee
where $p_{\omega1}$ ($p_{\omega3}$) is the $\omega$ momentum in the transition between the spin-single (spin-triplet) states, and 
\be
\Gamma[h_b(2P) \to h_b(1P) \, \pi \pi] \approx  \Gamma[\chi_{b1}(2P) \to \chi_{b1}(1P) \, \pi \pi]~. 
\label{gpp}
\ee
Indeed, both types of transitions are induced by the chromo-electric dipole interaction of the heavy quark pair with soft gluon field. The former transition arises in the third order in this interaction~\cite{mvomeg}, while the two-pion emission arises in the second order~\cite{gottfried}. In either process the heavy quark spin decouples, and the relation (\ref{go}) takes into account the difference in the phase space of the $S$-wave processes, which difference is quite essential, since the decay $\chi_{b1}(2P) \to \yp(1S) \, \omega$ is close to the threshold. In Eq.(\ref{gpp}) any kinematical difference can be neglected since the energy released in the two related decays is essentially the same within the (small) experimental uncertainty.

Using the relations (\ref{gg}), (\ref{go}) and (\ref{gpp}) one can readily find the estimates for the branching fractions for the yet unobserved hadronic transitions in terms of the experimentally measured quantities:
\bea
&&{\cal B}[h_b(2P) \to \eta_b(1S) \, \omega]= \nonumber \\
&&{ \Gamma[h_b(2P) \to \eta_b(1S) \, \omega] \over \Gamma[\chi_{b1}(2P) \to \yp(1S) \, \omega]} \, {\Gamma [\chi_{b1}(2P) \to \yp(2S) \, \gamma] \over \Gamma[h_b(2P) \to \eta_b(2S) \, \gamma]} \, {{\cal B}[\chi_{b1}(2P) \to \yp(1S) \, \omega] \over {\cal B}[\chi_{b1}(2P) \to \yp(2S) \, \gamma]} \,\times  \nonumber \\
&& {{\cal B}[h_b(2P) \to \eta_b(2S) \, \gamma]} \approx (7 \pm 2)\% ~,
\label{heo}
\eea
where the uncertainty is in fact dominated by the experimental errors in ${\cal B}[\chi_{b1}(2P) \to \yp(1S) \, \omega]$, and
\bea
&&{\cal B}[h_b(2P) \to h_b(1P) \, \pi \pi]= \nonumber \\
&&{ \Gamma[h_b(2P) \to h_b(1P) \, \pi \pi] \over \Gamma[\chi_{b1}(2P) \to \chi_{b1}(1P) \, \pi \pi]} \, {\Gamma [\chi_{b1}(2P) \to \yp(2S) \, \gamma] \over \Gamma[h_b(2P) \to \eta_b(2S) \, \gamma]} \, {{\cal B}[\chi_{b1}(2P) \to \chi_{b1}(1P) \, \pi \pi] \over {\cal B}[\chi_{b1}(2P) \to \yp(2S) \, \gamma]} \, \times  \nonumber \\
&& {{\cal B}[h_b(2P) \to \eta_b(2S) \, \gamma]} \approx (1.5 \pm 0.3)\% ~,
\label{hhpp}
\eea

With these estimates one can evaluate the balance of the total widths of the discussed bottomonium states and test whether the remaining fraction of the decays of the $h_b(2P)$ resonance can be made compatible with the expected similarity between the annihilation decay rates of the $P$-wave states. In doing so one can notice that for the lower $P$-wave levels the annihilation decay and the discussed radiative transitions $\chi_{b1}(1P) \to \yp(1S) \, \gamma$ and $h_b(1P) \to \eta_b(1S) \, \gamma$ exhaust the total probability of decay, modulo extremely minor decay modes like e.g. $\chi_{b1}(1P) \to \chi_{b0} \, \gamma$, or $h_b(1P) \to \yp(1S) \, \pi^0$, which can be safely neglected. For the $\chi_{b1}(2P)$ state the discussed hadronic transitions to lower bottomonium with emission of $\omega$ or two pions also contribute to the total decay rate. However, their total contribution is at the level of about two percent and can be readily taken into account (or neglected altogether). The counting is apparently somewhat different for the $h_b(2P)$ state. Indeed, as estimated in Eqs.~(\ref{heo}) and (\ref{hhpp}) the $\omega$ and two-pion transitions can contribute together about 8.5\% of the total decay rate, which is not negligible compared to the fraction remaining after accounting for the radiative decays. Furthermore we will argue that a potentially significant additional contribution can arise from the transition $h_b(2P) \to \yp(1S) \, \eta$, further reducing the estimated annihilation rate for the $h_b(2P)$ resonance.
We thus use here the following numbers for the annihilation branching fraction ${\cal B}_{ann}$ in the evaluation of the balance of decay rates of the discussed $P$-wave states:
\bea
&&{\cal B}_{ann} [\chi_{b1}(1P)] = 1- {\cal B}[ \chi_{b1}(1P) \to \yp(1S) \, \gamma]= (66.1 \, \pm 2.2)\%~; \nonumber \\
&&{\cal B}_{ann} [\chi_{b1}(2P)] = 1- {\cal B}[ \chi_{b1}(2P) \to \yp(1S) \, \gamma] -  {\cal B}[ \chi_{b1}(2P) \to \yp(2S) \, \gamma] \nonumber \\
&&-  {\cal B}[ \chi_{b1}(2P) \to \yp(1S) \, \omega] - {\cal B}[ \chi_{b1}(2P) \to \chi_{b1}(1P) \, \pi \pi] = (68.4 \pm 2.1)\%~; \nonumber \\
&&{\cal B}_{ann} [h_b(1P)] = 1- {\cal B}[ h_b(1P) \to \eta_b(1S) \, \gamma]= (50.8 \pm 8)\%~; \nonumber \\
&&{\cal B}_{ann} [h_b(2P)] + {\cal B} [h_b(2P) \to \yp(1S) \, \eta]= 1- {\cal B}[ h_b(2P) \to \eta_b(1S) \, \gamma]- {\cal B}[ h_b(2P) \to \eta_b(2S) \, \gamma] \nonumber \\
&&-{\cal B}[ h_b(2P) \to \eta_b(1S) \, \omega]- {\cal B}[ h_b(2P) \to h_b(1P) \, \pi \pi] \approx (22 \pm 15)\% ~.
\label{ann}
\eea

Using these numbers and the relations (\ref{gg}) between the radiative decay rates one can estimate the following `ratio of the ratios' of the absolute decay rates:
\be
r = {\left \{\Gamma_{ann} [h_b(2P)] + \Gamma[ h_b(2P) \to \yp(1S) \, \eta]\right \}/\Gamma_{ann} [h_b(1P)] \over \Gamma_{ann} [\chi_{b1}(2P)]/\Gamma_{ann} [\chi_{b1}(1P)]}~.
\label{rofr}
\ee 
If the similarity of the annihilation decay of the $P$-wave states holds, 
the quantity $r$ should be equal to one if the rate of the decay $h_b(2P) \to \yp(1S) \, \eta$ is negligible, and in general should be greater than one. The value of $r$ corresponding to the current data can be estimated by rewriting it as
\bea
&&r={\left \{ {\cal B}_{ann} [h_b(2P)] + {\cal B}[ h_b(2P) \to \yp(1S) \, \eta]\right \}/{\cal B}[ h_b(2P) \to \eta_b(2S) \, \gamma] \over {\cal B}_{ann} [h_b(1P)]/{\cal B}[ h_b(1P) \to \eta_b(1S) \, \gamma]} \, \times \nonumber \\
&& { {\cal B}_{ann} [\chi_{b1}(1P)]/{\cal B}[ \chi_{b1}(1P) \to \yp(1S) \, \gamma] \over {\cal B}_{ann} [\chi_{b1}(2P)]/ {\cal B}[ \chi_{b1}(2P) \to \yp(2S) \, \gamma]} \, \times \nonumber \\
&&{\Gamma[ h_b(2P) \to \eta_b(2S) \, \gamma] \over \Gamma [ \chi_{b1}(2P) \to \yp(2S) \, \gamma]} \, {\Gamma[ \chi_{b1}(1P) \to \yp(1S) \, \gamma] \over  \Gamma[ h_b(1P) \to \eta_b(1S) \, \gamma]} \approx 0.25 \pm 0.25 ~.
\label{rbg}
\eea
Clearly this estimate of $r$  shows that the current data on the radiative decays of $h_b(1P)$ and (especially of) $h_b(2P)$ are on the verge of a dramatic contradiction with the similarity of annihilation processes of the $P$-wave states of bottomonium. The disagreement may become even worse if the contribution of the decay $h_b(2P) \to \yp(1S) \, \eta$ in the last line in Eq.(\ref{ann}) is not small. 

Within the multipole expansion in QCD~\cite{gottfried,mv78} the transition $h_b(2P) \to \yp(1S) \, \eta$ arises as a combined effect of the chromoelectric dipole $(E1)$ and the chromomagnetic dipole $(M1)$ interaction described by the following terms in the effective Hamiltonian\be
H_{E1}=-{1 \over 2} \xi^a \, {\vec r} \cdot {\vec E}^a ~,
~~~H_{M1}= - {1 \over 2
\, m_b}\, \xi^a \, ({\vec \Delta} \cdot {\vec B}^a)~,
\label{cme}
\ee
where $\xi^a=t_1^a-t_2^a$ is the difference of the color generators
acting on the quark and antiquark (e.g. $t_1^a = \lambda^a/2$ with
$\lambda^a$ being the Gell-Mann matrices),  ${\vec r}$ is the vector
for relative position of the quark and the antiquark, ${\vec \Delta}=({\vec \sigma}_b
-{\vec \sigma}_{\bar b})/2$ is the
difference of the spin operators for the the quark and antiquark. Finally, ${\vec E}^a$ and $\vec B^a$ are the chromoelectric and
chromomagnetic components of the gluon field strength tensor. The assumed here normalization convention is that the QCD coupling $g$ is absorbed into the definition of the gluon field strength. 

The presence of the heavy quark mass in the denominator in $H_{M1}$ reflects the fact that the spin-dependent chromomagnetic interaction is suppressed by the heavy quark spin symmetry. In the considered process of emission of the $\eta$ meson this suppression however is somewhat compensated~\cite{vz,mv86} by the enhancement due to the axial anomaly relation~\cite{aanom1,aanom2}:
\be
\epsilon^{\mu \nu \lambda \sigma} \, \langle \eta | G_{\mu \nu}^a  G_{\lambda
\sigma}^a| 0
\rangle = 16 \pi^2 \, \sqrt{2 \over 3}
\, f_\eta \, m_\eta^2~,
\label{eanom}
\ee
where $f_\eta$ is the $\eta$ `decay
constant', equal to the pion decay constant $f_\pi \approx 130 \, MeV$
in the limit of exact flavor SU(3) symmetry, and $F_\eta$ is likely to
be larger due to effects of the SU(3) violation. 

The calculation of the transition rate is fully analogous to that for the $\yp(3S) \to h_b(1P) \, \pi^0$ decay in Ref.~\cite{mv86} (also in the review~\cite{mvc}), and the resulting expression can be written as
\be
\Gamma[h_b(2P) \to \yp(1S) \, \eta]= \left ( {\pi^2 \over 27} \, f_\eta \, m_\eta^2 \right )^2 \, |I(2P \to 1S)|^2 {p_\pi \over 3 \pi}~,
\label{hr}
\ee
where $p_\eta$ is the momentum of the $\eta$ meson and $I(2P \to 1S)$ is the heavy quarkonium matrix element:
\be
I(2P \to 1S) = {1 \over m_b} \, \langle 1S | {\cal G}_S \, r + r \, {\cal G}_P |2P \rangle
\label{isp}
\ee
containing the partial-wave Green function of the heavy quark pair ${\cal G}_S$ and ${\cal G}_P$ in the color octet state. 

Currently the matrix element (\ref{isp}) cannot be evaluated with any reliability, and one has to resort to indirect arguments. In particular, the rate of the discussed transition can be compared to that of a similar decay $\yp(3S) \to h_b(1P) \, \pi^0$. The latter decay involves isospin violation, which in terms of the chiral anomaly is expressed through the difference of the masses of the $u$ and $d$ quarks. The relation between the rates of these two processes takes the form
\bea
&&{\Gamma[h_b(2P) \to \yp(1S) \, \eta] \over \Gamma[\yp(3S) \to h_b(1P) \, \pi^0]}= {1 \over 3} \left ( {m_u+m_d \over m_u - m_d} \, {f_\eta \, m_\eta^2 \over f_\pi \, m_\pi^2 } \right )^2 \, {p_\eta \over p_\pi} \, \left | {I(2P \to 1S) \over I(3S \to 1P) } \right |^2 \approx \nonumber \\
&& 1.3 \times 10^3 \times \left | {I(2P \to 1S) \over I(3S \to 1P) } \right |^2~,
\label{ispr}
\eea
where the numerical value corresponds to $(m_d-m_u)/(m_d+m_u) = 0.3$ and $f_\eta=f_\pi$.

If the BaBar evidence~\cite{babar} for the decay $\yp(3S) \to h_b(1P) \, \pi^0$ is taken at face value, their signal corresponds to the absolute rate of this transition in the ballpark of 20\,eV. Thus if the matrix elements in Eq.(\ref{ispr}) for the $2P \to 1S$ and $3S \to 1P$ transitions were the same, the absolute rate of the decay $h_b(2P) \to \yp(1S) \, \eta$ would be about 25\,keV and would thus exceed the estimate~\cite{gr} ($\sim 15\,$keV) for the rate of the radiative transition $h_b(2P) \to \eta_b(2S) \, \gamma$. One can possibly argue, however, that the spatial size of the initial and the final bottomonium states in the transition $2P \to 1S$ is smaller than in $3S \to 1P$, so that the amplitude $I(2P \to 1S)$ should be somewhat suppressed as compared to $I(3S \to 1P)$ (although this argument does not take into account the possible effect of an extra oscillation in the $3S$ wave function). Allowing a factor of $\sim 1/2 \div 1/3$ for this suppression one can very approximately estimate the rate  $\Gamma[h_b(2P) \to \yp(1S) \, \eta]$ to be about one quarter of $\Gamma[h_b(2P) \to \eta_b(2S) \, \gamma]$ within a factor of two or so, corresponding to ${\cal B} [h_b(2P) \to \yp(1S) \, \eta] \sim O(10\%)$.

It is quite clear that the presented arguments involve a great uncertainty, and for this reason it would be very interesting if the transition $h_b(2P) \to \yp(1S) \, \eta$ could be found in the existing Belle data at the $\Upsilon(5S)$ energy, or an upper limit on the branching fraction for this process could be established. As is argued in this paper, the current data result in the estimate in Eq.(\ref{rbg}) which is in a really poor agreement with the `parton' picture of annihilation of the $P$-wave bottomonium, even if the contribution of this decay is negligible. An observation of the transition $h_b(2P) \to \yp(1S) \, \eta$ at a noticeable level would make the situation with the (non)similarity of the annihilation of the spin-singlet and spin-triplet $J=1$ bottomonium states quite dramatic and present an interesting riddle for theoretical interpretation.

The work of MBV is supported, in part, by the DOE grant DE-FG02-94ER40823.


\begin{thebibliography}{99}
\bibitem{belleh} 
  I.~Adachi {\it et al.}  [Belle Collaboration],
  Phys.\ Rev.\ Lett.\  {\bf 108}, 032001 (2012)
  [arXiv:1103.3419 [hep-ex]].
\bibitem{babar} 
  J.~P.~Lees {\it et al.}  [BABAR Collaboration],
  Phys.\ Rev.\ D {\bf 84}, 091101 (2011)
  [arXiv:1102.4565 [hep-ex]].
\bibitem{gr} 
  S.~Godfrey and J.~L.~Rosner,
  Phys.\ Rev.\ D {\bf 66}, 014012 (2002)
  [hep-ph/0205255].
\bibitem{bellehg} 
  R.~Mizuk {\it et al.}  [Belle Collaboration],
  arXiv:1205.6351 [hep-ex].
\bibitem{pdg}  
K. Nakamura {\it et. al.} [Particle Data Group], J.\ Phys.\ {\bf G37}, 075021 (2010) and 2011 partial update for the 2012 edition.  
\bibitem{mvomeg} 
  M.~B.~Voloshin,
  Mod.\ Phys.\ Lett.\ A {\bf 18}, 1067 (2003)
  [hep-ph/0304165].
\bibitem{gottfried}
  K.~Gottfried,
  Phys.\ Rev.\ Lett.\  {\bf 40}, 598 (1978).
\bibitem{mv78} 
  M.~B.~Voloshin,
  Nucl.\ Phys.\ B {\bf 154}, 365 (1979).
\bibitem{vz} 
  M.~B.~Voloshin and V.~I.~Zakharov,
  Phys.\ Rev.\ Lett.\  {\bf 45}, 688 (1980).
\bibitem{mv86} 
  M.~B.~Voloshin,
  Sov.\ J.\ Nucl.\ Phys.\  {\bf 43}, 1011 (1986)
  [Yad.\ Fiz.\  {\bf 43}, 1571 (1986)].
\bibitem{aanom1} 
  D.~J.~Gross, S.~B.~Treiman and F.~Wilczek,
  Phys.\ Rev.\ D {\bf 19}, 2188 (1979).
\bibitem{aanom2} 
  V.~A.~Novikov, M.~A.~Shifman, A.~I.~Vainshtein and V.~I.~Zakharov,
  Nucl.\ Phys.\ B {\bf 165}, 55 (1980).
\bibitem{mvc} 
  M.~B.~Voloshin,
  Prog.\ Part.\ Nucl.\ Phys.\  {\bf 61}, 455 (2008)
  [arXiv:0711.4556 [hep-ph]].

\end{thebibliography}
\end{document}